\documentclass[12pt]{article}

\usepackage{epsfig,amsfonts,amssymb,newlfont,rotate,float}

\DeclareMathAlphabet{\mathitbf}{T1}{cmr}{bx}{it}

\begin{document}

\title{Ising exponents in the two-dimensional site-diluted Ising model}
\author{H.~G.~Ballesteros\footnote{\tt hector@lattice.fis.ucm.es}~,~
        L.~A.~Fern\'andez\footnote{\tt laf@lattice.fis.ucm.es}~,\\
        V.~Mart\'{\i}n-Mayor\footnote{\tt victor@lattice.fis.ucm.es}~,~
        A.~Mu\~noz Sudupe\footnote{\tt sudupe@lattice.fis.ucm.es}~,\\
\normalsize \it Departamento de F\'{\i}sica Te\'orica I, 
        Facultad de CC. F\'{\i}sicas,\\  
\normalsize \it Universidad Complutense de Madrid, 28040 Madrid, Spain.\\
\\
        G.~Parisi\footnote{\tt giorgio.parisi@roma1.infn.it} ~and~
        J.~J.~Ruiz-Lorenzo\footnote{\tt ruiz@chimera.roma1.infn.it}~.\\
\normalsize \it Dipartimento di Fisica and Istituto Nazionale di 
        Fisica Nucleare,\\ 
\normalsize \it Universit\`a di Roma ``La Sapienza'', P.~A.~Moro  2, 
        00185 Roma, Italy.
}

\date{July 17, 1997}

\maketitle

\thispagestyle{empty}

\begin{abstract}
We study the site-diluted Ising model in two dimensions with Monte
Carlo simulations. Using finite-size scaling techniques we compute
the critical exponents observing deviations from the pure Ising ones.
The differences can be explained as the effects of
logarithmic corrections, without requiring to change the Universality Class.
\end{abstract}
  
\vskip 5 mm

Monte Carlo.
Disordered Systems.
Critical exponents.
Finite-size scaling.
$\epsilon$-expansion.


\newpage

Recently~\cite{KIM}, it has been reported using Monte Carlo
simulations, that the site-diluted Ising
model in two dimensions seems to present a second order transition
line with concentration dependent critical
indices. Previously, other authors~\cite{DOTSENKO,ANDREI} had claimed,
using analytical methods,
an Ising critical behavior corrected with logarithms.

The field theoretical predictions for this model (based on
Renormalization Group and Conformal Field Theory)~\cite{DOTPIPU}
assesses that disorder does not change the $\eta$ or $\nu$ exponents
since it only changes subleading terms. However, the specific heat,
at the critical point,
diverges as $\log(\log L)$ in the disordered system, while it does as
$\log L$ in the pure case $L$ being the lattice size.

In this letter, we extend the methods developed in ref.~\cite{4D} for
the four dimensional site-diluted Ising model to the two-dimensional
case.

We observe that, although an apparent variation of the indices seems
to happen when varying the concentration, this can be explained as
a transient effect. In fact, a pure Ising value for the indices plus
logarithmic corrections fits very well our Monte Carlo (MC) data.
When preparing this letter, other authors have also reported similar
conclusions in the bond diluted Ising model using different 
techniques~\cite{2Dbis}.

The numerical methods as well as the analytical computations based
on the perturbative renormalization group are very similar in four
and two dimensions, so we will report here just the minimal details
required, centering ourselves in the description of the results.
We address to reference~\cite{4D} for further details of the method.

Our procedure is based on a Finite-Size Scaling (FSS) analysis. We perform
MC simulations in the critical region for several values of
the concentration $p$. For each $p$ value, we generate hole
configurations in a square lattice filling the sites randomly with
probability $p$. 

For each hole configuration (sample), we perform a MC simulation of
the Ising model defined as the set of spins lying in the filled sites
coupled through a nearest-neighbor interaction.  In the smallest
concentration we use the cluster
Swendsen-Wang algorithm~\cite{SW} to update the signs of the
spins. For the rest of concentrations, we found the Wolff single cluster
version of this algorithm~\cite{WOLFF} to be more efficient. At each
sample, we measure the energy, the magnetization and the Fourier
transform of the Ising field at minimal momenta in about 100
independent spin configurations. We store the independent measures in
order to compute derivatives with respect to the coupling and to
extrapolate the results to close values of the coupling and the
dilution. We have extrapolated in $\beta$ in all cases but $p\simeq
2/3$ where a $p$ extrapolation performs better.

To reduce statistical errors, we have generated $10,000$ hole
samples for each $(L,p)$ pair, $L$ being the lattice size. We
have simulated at concentrations: $p=1,8/9,3/5,2/3$.  Let us
remember that the percolation threshold is at $p_\mathrm{c}\simeq 0.59$
~\cite{STAUFFER}.
The FSS method that we use~\cite{OURFSS} is based upon the
ratio of several observables: magnetization, susceptibility, correlation
length, Binder cumulants, and their derivatives for two different lattice
sizes $L_1$ and $L_2$. At the parameter values where the correlation
lengths ratio matches $L_2/L_1$ we expect that, in the absence of 
logarithmic corrections
\begin{equation}
\frac{O(L_2,\beta,p)}{O(L_1,\beta,p)}=(L_2/L_1)^{x_O/\nu}+
O(L_1^{-\omega})\ ,
\label{FSS}
\end{equation}
where $x_O$ is the critical exponent for the observable $O$,
e.g. $\gamma$ for the susceptibility, $\nu$ for the correlation
length, etc. $\omega$ is the universal corrections-to-scaling exponent.
The $\beta$-derivatives of the previous observables go
with the corresponding exponents plus $1/\nu$.  In all cases we use
pairs of lattices of sizes $L$ and $2L$.

\begin{table}[t]
\footnotesize
\begin{center}
\begin{tabular}{|r|l|l|l|l|}\hline
$L$&   \multicolumn{1}{c|}{$p=1.0$}      
    & \multicolumn{1}{c|}{$p=0.88889$}      
    & \multicolumn{1}{c|}{$p=0.75$}      
    & \multicolumn{1}{c|}{$p=0.6666$}   \\ \hline\hline
24& 1.009(4) &&&\\\hline
32& 1.009(5) &1.078(6)&1.142(9)&1.182(12)\\\hline
48& 0.995(4) &1.072(6)&1.114(9)&1.170(12)\\\hline
64& 1.001(4) &1.079(5)&1.112(9)&1.151(10)\\\hline
96& 1.010(5) &1.075(6)&1.111(9)&1.125(12)\\\hline
128&0.993(4) &1.066(5)&1.098(8)&1.141(13)\\\hline
192& &1.065(6)&1.096(9)&1.140(13)\\\hline
256&1.004(5) &&&\\\hline
\end{tabular}
\caption{The $\nu$ exponent for $(L,2L)$ pairs at different
dilutions.}
\label{NUNAIVE}
\end{center}
\end{table}

\begin{table}[t]
\footnotesize
\begin{center}
\begin{tabular}{|r|l|l|l|l|}\hline
$L$&  \multicolumn{1}{c|}{$p=1.0$}      
    & \multicolumn{1}{c|}{$p=0.88889$}      
    & \multicolumn{1}{c|}{$p=0.75$}      
    & \multicolumn{1}{c|}{$p=0.6666$}   \\ \hline\hline
24&0.2465(7)  &&&\\\hline
32&0.2466(8)  &0.2495(9)&0.2504(10)&0.2454(18)\\\hline
48&0.2495(8)  &0.2487(7)&0.2498(11)&0.2448(14)\\\hline
64&0.2499(8)  &0.2490(8)&0.2469(8)&0.2455(15)\\\hline
96&0.2498(8)  &0.2501(8)&0.2460(9)&0.2456(14)\\\hline
128&0.2497(8) &0.2495(8)&0.2469(9)&0.2461(13)\\\hline
192&            &0.2485(8)&0.2483(7)&0.2483(14)\\\hline
256&0.2517(7) &&&\\\hline
\end{tabular}
\caption{The $\eta$ exponent for $(L,2L)$ pairs at different
dilutions.}
\label{ETANAIVE}
\end{center}
\end{table}

In table 1 and 2 we report the results for the exponents $\nu$ and
$\eta$ respectively, using the relation (\ref{FSS}) for the
$\beta$-derivative of the correlation length in the former case and
the susceptibility in the latter (after applying the scaling relation
$\eta=2-\gamma/\nu$).  Notice that there is some statistical
anticorrelation between sucessive even (odd) rows.  We recall the critical
exponents in the $\beta\to\infty$ limit (pure site percolation),
$\nu=4/3$ and $\eta=5/24$ conjectured by Nienhuis~\cite{NIEN} and in
agreement with a recent MC work~\cite{PERC}.

Although the $\nu$ exponent seems to be non-constant as a function of
the concentration, the values for $\eta$ are very stable. This fact
was also observed in  the scaling of the Yang-Lee zeroes at the
critical point~\cite{JJRL}.

An interpretation of these results could be a continuous set of
Universality Classes, as assumed in reference~\cite{KIM}.
However, a simpler scenario is a fixed value for $\nu$ (that
of the Ising Model) plus logarithmic corrections.

Using the results from refs.~\cite{DOTSENKO,SHALAEV}
it is possible to show that the
derivative of the correlation length with $\beta$ 
reads at critical point as
\begin{equation}
\partial_\beta \xi \propto -\frac{\xi^2}{\sqrt{1+C \log L}}
\left( {1+\frac{C}{2+2C \log L}} \right)
\label{corre}
\end{equation}
where $C$ depends on the dilution. Computing the $\nu$ exponent from 
(\ref{FSS}) and (\ref{corre}) 
we obtain an apparent behavior as
\begin{equation}
\nu^\mathrm{apparent}=1+A/\log L+\ldots
\label{LOG}
\end{equation}

\begin{figure}[t]
\begin{center}
\leavevmode
\centering\epsfig{file=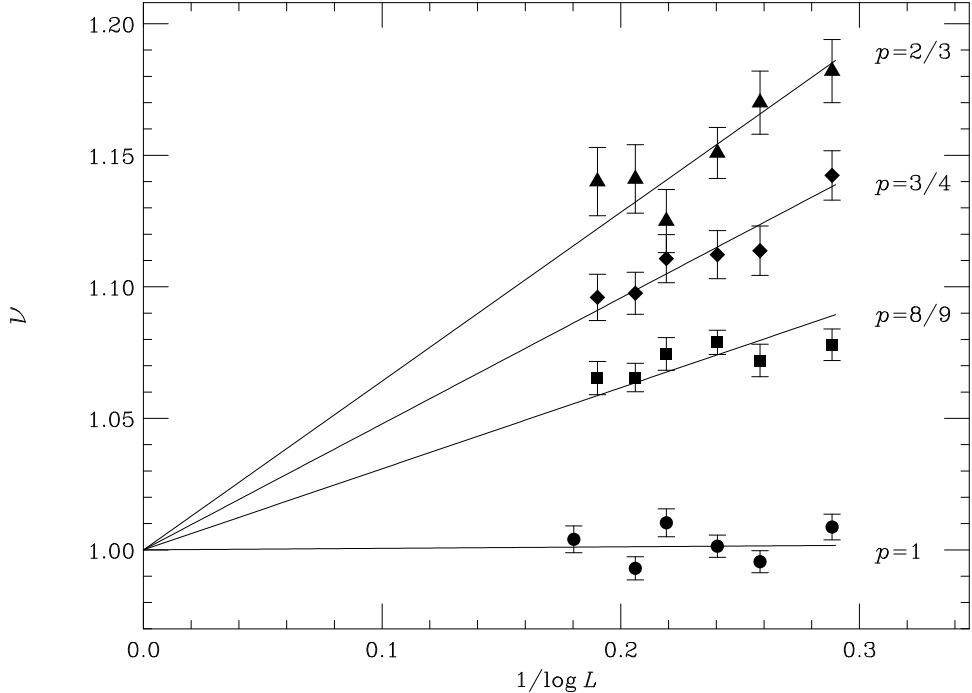,width=0.68\linewidth,angle=90}
\end{center}
\protect\caption{$\nu$ exponent as a function of $1/\log L$ for
several values of $p$.
\protect\label{NULOG}}
\end{figure}

In figure 1 we plot the values of $\nu$ reported in table 2 as a
function of $1/\log L$. All our MC data are compatible with a
behavior of type (\ref{LOG}).

\begin{table}[t]
\footnotesize
\begin{center}
\begin{tabular}{|l|l|l|c|c|}\hline
      \multicolumn{1}{|c|}{$p_\mathrm{c}$}      
    & \multicolumn{1}{c|}{$\beta_\mathrm{c}$}      
    & \multicolumn{1}{c|}{$\omega$}      
    & \multicolumn{1}{c|}{$\chi^2/\mathrm{d.o.f.}$}      
    & \multicolumn{1}{c|}{$L_1-(L_2^a-L_2^b)$}\\ \hline\hline
1            & 0.440682(5) & 1.5(8)   &15.2/12 &24-(32-512)\\\hline
0.88889      & 0.53781(2)  & 1.0(3)   & 2.6/8  &48-(64-384)\\\hline
0.75         & 0.77125(8)  & 0.9(5)   & 7.7/6  &48-(96-384)\\\hline
0.66661(3)   & 1.10        & 0.6(2)   & 9.6/8  &32-(64-384)\\\hline
\end{tabular}
\caption{Critical parameters for several dilutions. We also show the
computed corrections to scaling exponent $\omega$, the fit quality and
the range of lattice sizes used (crossings from $L_1-L_2^a$ to $L_1-L_2^b$).}
\label{BETAC}
\end{center}
\end{table}

The critical parameters can be computed with great accuracy studying
the crossing of several quantities (second momentum correlation length 
divided by the lattice length, and Binder parameter for the
magnetization) for the different lattice sizes. The deviation of
the crossing point for lattices sizes $L,sL$ scales as
\begin{equation}
\Delta \beta_\mathrm{c}^L, \Delta p_\mathrm{c}^L
\propto \frac{1-s^{-\omega}}{s^{1/\nu}-1}L^{-\omega-1/\nu}\ ,
\end{equation}
We address to reference~\cite{OURFSS} for details of the method. The
results are reported in table~\ref{BETAC}. 
Notice that for the Ising
Model the method gives the correct value within errors for the
critical temperature (1 part in
$10^5$). Regarding the $\omega$ exponent, whose conjectured value is
$\omega=4/3$~\cite{ZINN-JUSTIN} for the pure system, 
to obtain more accurate results, simulations on smaller lattices 
should be added, but it
would increase the systematic errors in the determination of the
critical coupling.

\begin{figure}[t]
\begin{center}
\leavevmode
\centering\epsfig{file=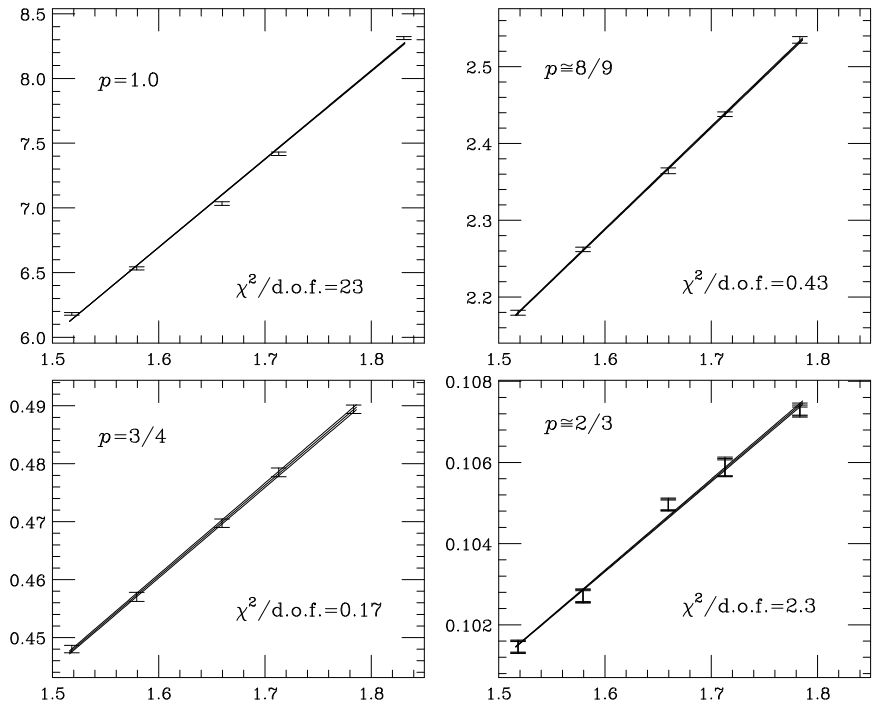,width=0.68\linewidth,angle=90}
\end{center}
\protect\caption{Specific heat as a function of the double logarithm of
the lattice size. The lines correspond to linear fits for $L\ge 96$.
The three lines for each case show the effect of the error in the
determination of the critical parameters.
\protect\label{CV}}
\end{figure}

We finally measure the specific heat $C_V$ at the critical points.
The results are displayed in figure~\ref{CV}. Excluding the Ising
limit, a linear function of $\log(\log L)$, as predicted in
ref.~\cite{DOTSENKO}, fits very well the data.

We thus conclude that the site diluted Ising model in two dimensions belongs
to the same Universality Class as the pure Ising model, although
we find strong logarithmic effects.

We thank to the CICyT (contracts AEN94-0218, AEN96-1634) for partial
financial support. JJRL is granted by EC HMC (ERBFMBICT950429).

\end{document}